\documentclass[pdflatex,sn-mathphys-num]{sn-jnl}

\usepackage{graphicx}%
\usepackage{multirow}%
\usepackage{amsmath,amssymb,amsfonts}%
\usepackage{amsthm}%
\usepackage{mathrsfs}%
\usepackage[title]{appendix}%
\usepackage{xcolor}%
\usepackage{textcomp}%
\usepackage{manyfoot}%
\usepackage{booktabs}%
\usepackage{algorithm}%
\usepackage{algorithmicx}%
\usepackage{algpseudocode}%
\usepackage{listings}%

\theoremstyle{thmstyleone}%
\theoremstyle{thmstyletwo}%

\theoremstyle{thmstylethree}%

\graphicspath{ {./} }

\begin{document}

\title[Article Title]{Transfer Dynamics and Spectral Cascades in Graph-Coupled Kuramoto Networks}


\author[1]{\fnm{Marcin} \sur{Kowalczyk} } \email{mt.kowalczyk3@student.uw.edu.pl}
\author[2]{\fnm{Pietro} \sur{Li\`{o}}} \email{pl219@cam.ac.uk}
\author*[1,3]{\fnm{Zbigniew} \sur{Struzik} } \email{z.r.struzik@p.u-tokyo.ac.jp}

\affil[1]{\orgdiv{Faculty of Physics}, \orgname{University of Warsaw}, \orgaddress{\street{Pasteura 5}, \postcode{02-093}, \city{Warsaw}, \country{Poland}}}
\affil[2]{\orgdiv{Department of Computer Science and Technology}, \orgname{University of Cambridge}, \orgaddress{\street{The Old Schools, Trinity Lane}, \city{Cambridge}, \postcode{CB2 1TN} \country{UK}}}
\affil[3]{\orgdiv{Graduate School of Education}, \orgname{University of Tokyo}, \orgaddress{\street{7‑3‑1 Hongo, Bunkyo‑ku}, \city{Tokyo}, \postcode{113‑0033}, \country{Japan}}}


\abstract{This work introduces a spectral transfer formalism for nonlinear synchronisation dynamics to resolve how internal dynamical activity redistributes across network scales, moving beyond traditional global observables like the Kuramoto order parameter. By projecting Kuramoto phase dynamics onto the eigenbasis of the graph Laplacian, we construct a time-dependent transfer matrix and derive novel network metrics, including the spectral flux, which quantifies directional redistribution between structural dynamical scales rather than physical transport across nodes. Numerical simulations on modular and hierarchical networks reveal highly structured, intermittent spectral transfer episodes characterised by forward and inverse spectral cascades, directional reversals, and complex interaction geometries. Crucially, we demonstrate a clear separation between macroscopic coherence and microscopic spectral interactions, showing that highly organised, fluctuating internal redistribution dynamics persist even beneath nearly stationary and featureless global synchronisation states. This framework extends conventional synchronisation analysis and establishes a multiscale approach for interpreting network dynamics as an evolving process of spectral reorganisation.}


\maketitle

\section{Introduction}\label{sec1}

Synchronisation phenomena emerge across a remarkably broad range of natural and engineered systems, including neuronal networks, power grids, chemical oscillators, ecological systems, and social dynamics. The collective organisation of interacting oscillatory units has therefore become one of the central themes of nonlinear science and network dynamics.

A major theoretical framework for studying synchronisation processes is provided by coupled phase oscillator models, among which the Kuramoto model \cite{Kuramoto1975, Kuramoto1984} occupies a particularly important role due to its conceptual simplicity and rich dynamical behaviour. Over the past decades, extensive research has investigated synchronisation transitions \cite{Acebron2005}, phase coherence, cluster formation, chimera states, explosive synchronisation, and the influence of network topology on collective dynamics \cite{Barahona2002, Pecora1998}.

At the same time, increasing attention has been devoted to the role of graph structure in shaping dynamical processes on networks. Spectral methods based on graph Laplacians provide a natural multiscale representation of network organisation, allowing collective dynamics to be analysed in terms of structural modes associated with different graph scales. In synchronisation studies, Laplacian eigenmodes have previously been used to characterise stability, synchronisation pathways, and collective organisation in complex networks \cite{Arenas2008,Rodrigues2016}.

Despite these advances, synchronisation dynamics are still most commonly characterised through global observables such as coherence or synchronisation order parameters \cite{Pecora1998}. While these quantities successfully quantify the degree of collective organisation, they provide only limited information about how dynamical activity redistributes internally across graph scales during the synchronisation process.

This raises a natural and largely unexplored question: Can synchronisation dynamics exhibit structured transfer processes between graph spectral modes \cite{Chung1997, Newman2018} analogous to cascade-like \cite{Frisch1995, Kraichnan1967} redistribution across dynamical scales?
Addressing this question requires moving beyond purely macroscopic descriptions of synchronisation and developing a framework capable of resolving the internal spectral organisation of network dynamics.

In this work, we introduce a spectral transfer formalism for nonlinear synchronisation dynamics on networks. By projecting the dynamics onto the eigenbasis of the graph Laplacian, we construct a time-dependent transfer matrix describing nonlinear interactions between graph spectral modes. This framework allows synchronisation dynamics to be interpreted as an evolving redistribution process across graph structural scales.
Using this approach, we demonstrate the emergence of several nontrivial phenomena such as: intermittent spectral transfer, spectral cascades and strongly fluctuating internal redistribution dynamics occurring beneath nearly stationary global synchronisation states.

A central observation of the present work is that conventional synchronisation observables may remain comparatively smooth and featureless even while the internal spectral transfer dynamics undergo rapid and highly structured reorganisations. The resulting behaviour reveals a separation between macroscopic coherence and microscopic spectral interaction organisation.

To characterise this regime, we introduce a set of observables quantifying modal persistence, interaction-mediated transfer and spectral flux. The analysis reveals that synchronisation dynamics can self-organise into temporally localised transfer episodes involving strong redistribution between graph spectral modes. These episodes exhibit intermittent burst structure, directional reversals, and complex interaction geometries that are largely hidden from conventional order-parameter descriptions.

Importantly, the spectral flux introduced here does not represent physical transport on the underlying graph. Rather, it quantifies redistribution between structural dynamical scales defined by the graph Laplacian eigenbasis. The resulting cascade-like processes therefore correspond to evolving patterns of spectral organisation rather than motion of conserved quantities across physical space.

The present study focuses primarily on modular network architectures, which naturally support multiscale organisation and competing synchronisation pathways. However, the proposed framework is considerably more general and may potentially be extended to broader classes of nonlinear network dynamics.

The paper is organised as follows. In Section II we introduce the spectral transfer framework and define the modal interaction observables. Section III presents the computational methodology. Section IV investigates the dependence of the spectral transfer cascades and intermittent transfer episode on coupling strength and network parameters. Finally, Section V discusses the broader implications and possible extensions of the proposed framework.

\section{Theoretical Framework}\label{sec2}

In this section we introduce the spectral framework used to characterise nonlinear transfer dynamics in synchronising network systems. The central idea is to project the network dynamics onto the eigenmodes of the graph Laplacian and to study how dynamical activity redistributes between these modes over time. This approach allows synchronisation dynamics to be interpreted not only as the emergence of global coherence, but also as a process of evolving spectral organisation involving persistent modes, inter-modal interactions, and cascade-like transfer events.

\subsection{Modular Networks and Network Dynamics}

We begin by specifying the class of networks considered in this study. Our objective is to investigate spectral transfer processes in systems that exhibit meaningful multiscale organisation. Random graphs often possess relatively homogeneous structural properties, which can limit the emergence of strongly differentiated transfer patterns across graph scales.

To investigate non-trivial spectral transfer dynamics, we therefore focus on modular networks. Modular networks consist of densely connected communities linked by comparatively sparse inter-community connections. Such architectures introduce persistent structural heterogeneity and naturally support organisation across multiple graph scales, providing a suitable setting for the study of graph-spectral transfer processes.

Modular networks arise in a wide range of natural and engineered systems, including social networks, biological networks, neural systems, and information networks such as the World Wide Web. From the perspective of transfer dynamics, they are particularly attractive because interactions within densely connected communities may differ substantially from interactions mediated by the relatively sparse links connecting different communities. This separation of scales creates conditions under which non-trivial redistribution processes can emerge in the graph-spectral domain.

\begin{figure}[h]
\centering
\includegraphics[width=\textwidth]{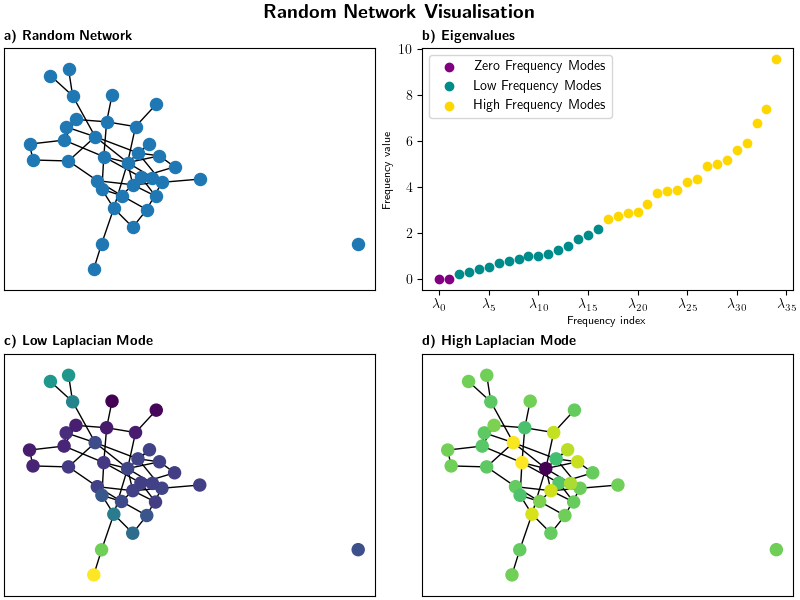}
\caption{Illustration of graph spectral decomposition. (a) Example of a random network. (b) Laplacian spectrum of the network shown in panel (a), with graph frequencies ordered by eigenvalue. (c) First non-trivial Laplacian eigenvector of the same network, representing a low-frequency graph mode associated with large-scale structural organisation. (d) Highest-frequency Laplacian eigenvector of the network, representing fine-scale variation across the graph. Together, panels (b)--(d) illustrate how the network structure shown in panel (a) gives rise to a hierarchy of graph-spectral modes that form the basis of the graph Fourier representation used throughout this work.}
\label{fig:RandomGraph}
\end{figure}

For illustration, panel \textbf{a)} of Fig.~\ref{fig:RandomGraph} shows a representative random network, while panel \textbf{a)} of Fig.~\ref{fig:ModularGraph} shows a modular network with clearly identifiable community structure.

Having specified the network architecture, we next introduce the dynamical process evolving on the graph. Since the objective of this work is not to propose a new oscillator model but rather to analyse dynamical behaviour from a graph-spectral perspective, we employ the well-established Kuramoto model \cite{Kuramoto1975}. Owing to its simplicity and rich nonlinear behaviour, the Kuramoto model has become one of the standard frameworks for studying synchronisation phenomena in complex networks \cite{Kuramoto1984,Acebron2005,Rodrigues2016}.

Accordingly, we consider a modular network of $N$ coupled oscillators evolving according to the Kuramoto dynamics:

\begin{equation}
    \dot{\theta}_i = \omega_i + K \sum_{j=1}^{N} A_{ij} \sin(\theta_j-\theta_i), \qquad i=1,\dots,N ,
    \label{eq:Kuramoto}
\end{equation}
where:
\begin{itemize}
    \item $\theta_i(t)$ denotes the phase of node i,
    \item $\dot{\theta}_i$ denotes the phase change of node i,
    \item $\omega_i$ is its natural frequency,
    \item $K$ is the global coupling strength,
    \item $A_{ij}$ is the adjacency matrix of the network.
\end{itemize}

Given the initial conditions, equation \eqref{eq:Kuramoto} is numerically integrated to obtain the phase evolution of all oscillators. The resulting time-dependent phase configuration defines a graph signal that can subsequently be analysed in the graph-spectral domain introduced in the following section.


\subsection{Graph Spectral Decomposition and Spectral Modal Energy}
A network consisting of nodes and edges can be represented through its graph Laplacian \cite{Sandryhaila2013,Ortega2018}.
It is defined as $L=D-A$, where $D$ is the degree matrix and $A$ is the adjacency matrix. Since $L$ is symmetric for undirected graphs, it possesses an orthonormal basis of eigenvectors
\begin{equation}
L\phi_m=\lambda_m\phi_m, \qquad m=0,\dots,N-1 ,
\end{equation}
with eigenvalues ordered as
\begin{equation}
0=\lambda_0\le \lambda_1\le\dots\le\lambda_{N-1} .
\end{equation}

\begin{figure}[h]
\centering
\includegraphics[width=\textwidth]{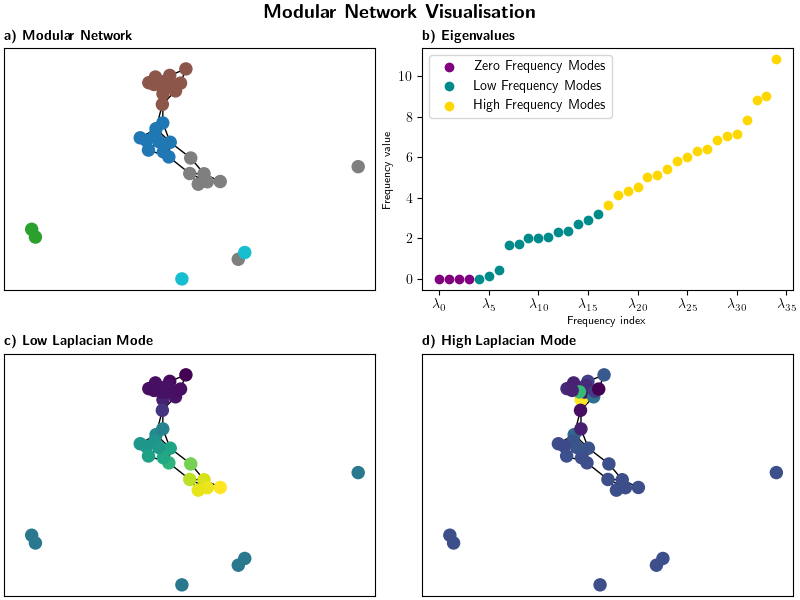}
\caption{Illustration of the modular network architecture used throughout this study. (a) Example of a modular network, with node colours indicating community affiliation. (b) Laplacian spectrum of the network shown in panel (a), with graph frequencies ordered by eigenvalue. (c) First non-zero Laplacian eigenvector of the same network, representing a low-frequency mode associated with large-scale community structure. (d) Highest-frequency Laplacian eigenvector of the network, representing fine-scale variation across the graph. Together, panels (b)--(d) illustrate how the modular structure shown in panel (a) is encoded in the graph-spectral representation used throughout this work.}
\label{fig:ModularGraph}
\end{figure}

The Laplacian eigenvectors define natural structural modes of the network \cite{Chung1997, Shuman2013}. Low-frequency modes typically correspond to large-scale collective organisation, whereas high-frequency modes capture finer structural variations and localised patterns. Figures \ref{fig:RandomGraph}c, \ref{fig:RandomGraph}d  and Figures \ref{fig:ModularGraph}c, \ref{fig:ModularGraph}d illustrate representative examples of such modes.
These modes can be used to transform graph signals to spectral domain by graph fourier transform:
\begin{equation}
a_m(t) = \sum_{i=1}^{N} \theta_i(t)\phi_m^T(i) ,
\label{eq:GFT}
\end{equation}
where $a_m(t)$ is the amplitude of mode $m$.
The transform can be reversed by inverse graph fourier transform
\begin{equation}
\theta_i(t) = \sum_{m=0}^{N-1} a_m(t)\phi_m(i).
\label{eq:IGFT}
\end{equation}

To characterise the distribution of dynamical activity across graph scales, we define the spectral modal energy
\begin{equation}
E_m(t)=|a_m(t)|^2.
\label{eq:SpectralEnergy}
\end{equation}
This quantity measures the contribution of mode $m$ to the network state. The evolution of $E_m(t)$ reveals how synchronisation activity redistributes spectrally during the dynamics. In particular how concentration in low modes corresponds to large-scale coherent organisation and how activation of higher modes corresponds to finer spectral structure and increased multiscale complexity.
This representation leads to cascade-like transfer processes between graph spectral scales, which can be seen in Figure \ref{fig:ModalVariables}a.

\subsection{Transfer Matrix}
\label{sec:transfer_matrix}
The graph Fourier decomposition introduced in the previous section provides a representation of the network dynamics in terms of graph spectral modes. The corresponding modal amplitudes describe how dynamical activity is distributed across graph scales at a given instant. However, the modal amplitudes alone do not reveal how activity is exchanged between modes through the nonlinear dynamics.

Two network states may exhibit similar modal amplitude distributions while differing substantially in the interactions responsible for maintaining those distributions. In other words, the graph Fourier decomposition describes \emph{where} activity is located in graph spectral space, but does not explain \emph{how} activity is redistributed between graph modes. To address this question, we introduce a transfer matrix formalism that quantifies interactions among graph Fourier modes.
Consider the graph Fourier decomposition
\begin{equation}
\theta_i(t) = \sum_{m=1}^{N} a_m(t)\phi_m(i),
\end{equation}
where $\phi_m(i)$ denotes the $m$-th Laplacian eigenvector and $a_m(t)$ is its corresponding modal amplitude. 

The set of modal amplitudes provides a compact multiscale representation of the network state. The nonlinear coupling term of the Kuramoto model, however, induces interactions between these modes, causing activity initially associated with one graph scale to influence others.
The objective of the transfer matrix is to quantify these interactions. For each source mode $m$, its contribution is first reconstructed in the node domain using the corresponding graph Fourier component. The reconstructed mode is subsequently propagated through the nonlinear coupling term of the Kuramoto dynamics, thereby isolating the influence of that mode on the instantaneous evolution of the network. The resulting nonlinear forcing field is then projected back onto the complete graph Fourier basis.
This procedure yields a set of coefficients describing how activity originating from source mode $m$ contributes to every target mode $k$. The corresponding transfer matrix element
\begin{equation}
T_{m \rightarrow k}(t)
\end{equation}
quantifies the instantaneous transfer from mode $m$ to mode $k$.

\subsubsection{Transfer Matrix Construction}
Having introduced the transfer matrix, let us now consider the computational aspects.
Let us assume that the Laplacian eigenvectors $\phi$ and modal amplitudes $a(t)$ corresponding to graph signal created by Kuramoto dynamics have been computed.
To quantify interactions originating from mode $m$, the corresponding modal contribution is reconstructed in the spatial domain
\begin{equation}
\theta^{(m)}_i(t) = a_m(t)\phi_m(i),
\end{equation}
where $\theta^{(m)}_i(t)$ denotes the graph signal on $i$-th node from $m$-th Laplacian mode. 
This reconstructed signal is then inserted into the nonlinear coupling term of the Kuramoto dynamics, allowing the influence of mode $m$ on the network evolution to be isolated
\begin{equation*}
    F_m(i,t) = K \sum_{j=1}^N A_{ij} sin(\theta_j^{(m)} - \theta_i^{(m)}).
\end{equation*}
The resulting nonlinear forcing field is subsequently projected back onto the complete graph Fourier basis 
\begin{equation*}
   f_{m\to k}(t) = \sum_{i=1}^N F_m(i,t) \phi_k(i).
\end{equation*}
This projection yields a set of coefficients describing how the activity associated with mode $m$ contributes to all target modes $k$.
To quantify modal interactions, we introduce a transfer matrix $T_{m \to k }(t)$
\begin{equation*}
   T_{m \to k}(t) = 2 a_k(t)\, f_{m\to k}(t),
\end{equation*}
whose elements measure the contribution associated with mode $m$ to the evolution of mode $k$.

\begin{figure}[H]
\centering
\includegraphics[width=\textwidth]{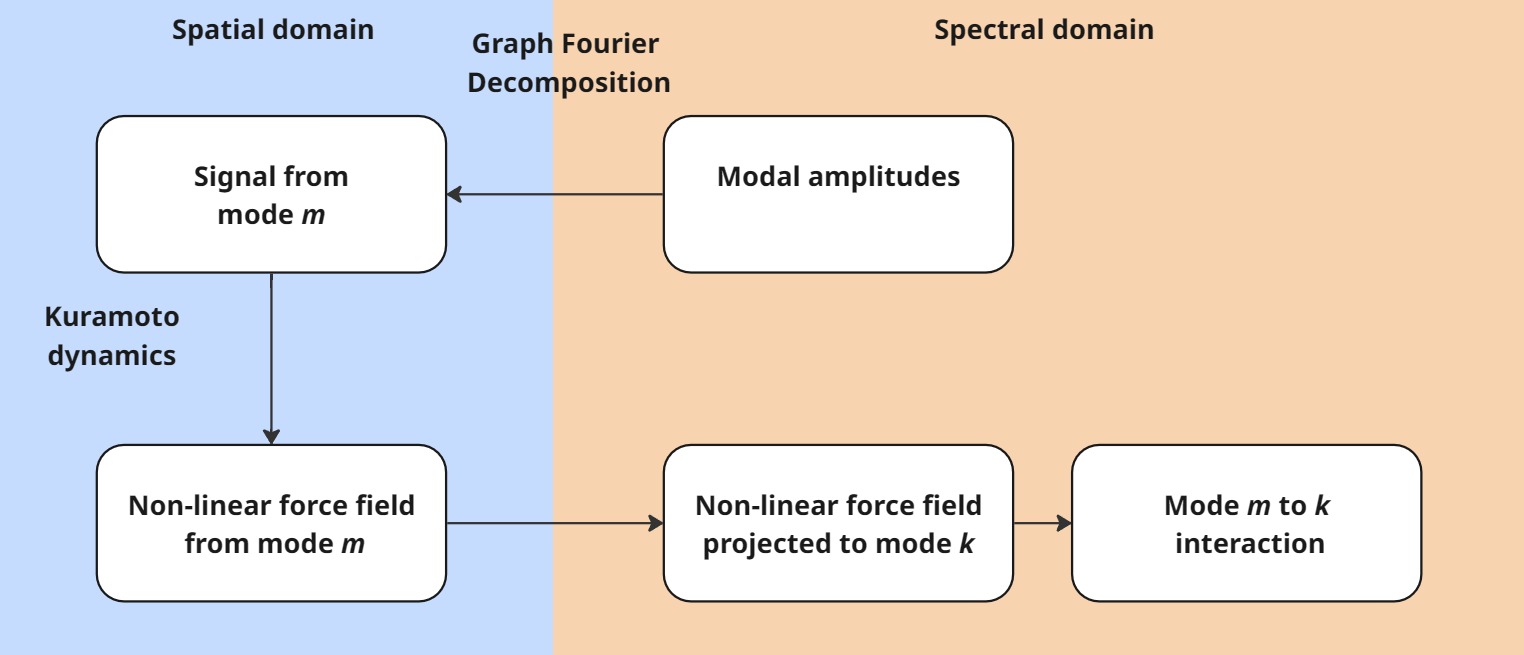}
\caption{Transfer matrix construction schematic display steps described in Section \ref{sec:transfer_matrix}}
\label{fig:transfer_schematic}
\end{figure}

\subsubsection{Transfer Matrix Interpretation}
The matrix may be interpreted as a graph-spectral interaction network whose nodes correspond to graph Fourier modes and whose directed edges represent instantaneous transfers between graph scales. The diagonal elements describe modal persistence and self-interaction, whereas the off-diagonal elements quantify exchanges between distinct graph modes. Strong off-diagonal structures therefore indicate active redistribution of dynamical activity throughout graph spectral space.
A schematic representation of the transfer matrix construction is shown in Figure \ref{fig:transfer_schematic}. Starting from the network state, the dynamics are decomposed into graph Fourier modes, mode-specific contributions are isolated and propagated through the nonlinear coupling term, and the resulting interactions are projected back onto the spectral basis. The resulting transfer matrix provides a compact representation of graph-spectral redistribution processes.
%

\begin{figure}
\centering
\includegraphics[width=0.9\textwidth]{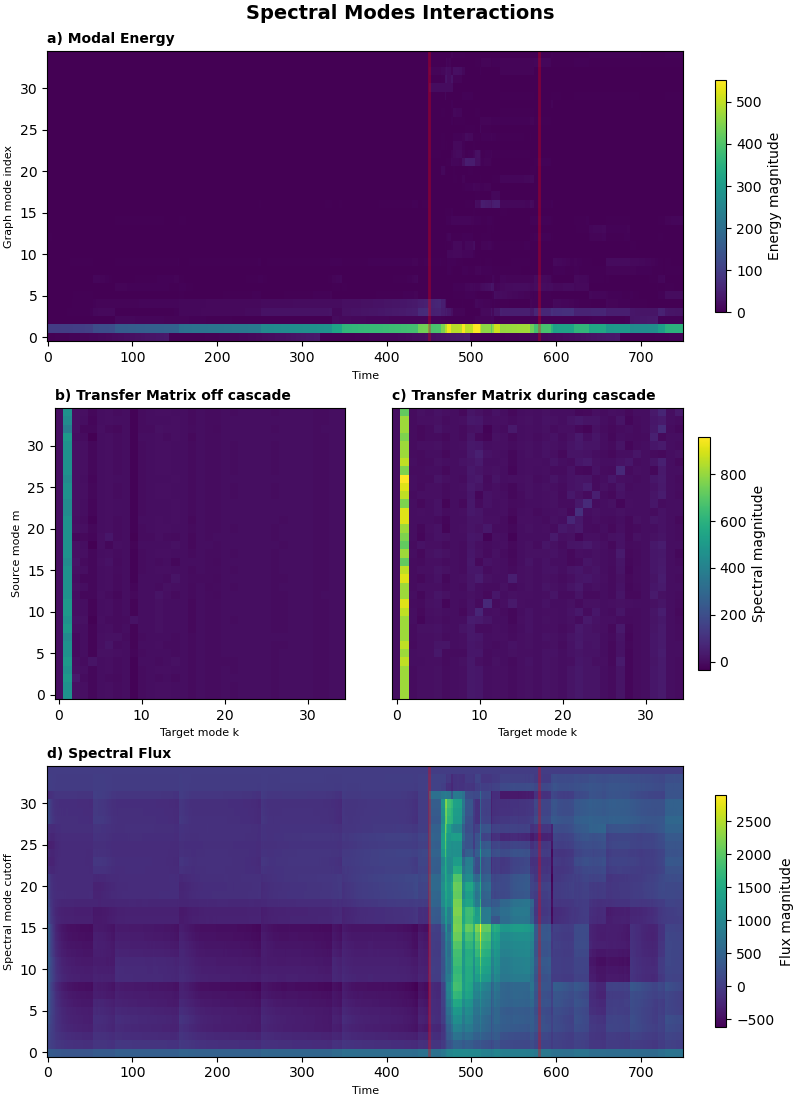}
\caption{Emergence of graph-spectral cascade dynamics. (a) Temporal evolution of the modal energy $E_m(t)$, showing persistent low-frequency organisation together with intermittent redistribution across higher graph-spectral modes. Cross-sections of the transfer matrix $T_{m\to k}(t)$ are shown (b) outside a cascade episode and (c) during an active cascade event. Panel (d) shows the corresponding spectral flux $\Pi(t,K_0)$. Red vertical lines in panels (a) and (d) indicate a representative cascade episode.
}
\label{fig:ModalVariables}
\end{figure}

Representative cross-sections of the transfer matrix are shown in Figure~\ref{fig:ModalVariables}. Figure~\ref{fig:ModalVariables}b illustrates the interaction structure outside a cascade episode, where most modes contribute predominantly to the first non-zero mode. Figure~\ref{fig:ModalVariables}c shows the interaction structure during a cascade event. While the first non-zero mode remains an important target of transfer, stronger diagonal contributions and increased interactions involving higher modes become visible.
The transfer matrix constitutes the central object of the present framework and serves as the basis for all transfer observables introduced in the following section. In particular, spectral flux, transfer volatility, occupancy statistics, reversal rates, and burstiness measures are all derived from the temporal evolution of the transfer matrix and provide complementary descriptions of graph-spectral transfer dynamics.

\subsubsection{Spectral Flux and Cascade Directionality}

To characterise the directionality of spectral redistribution, we define a net spectral flux
\begin{equation}
    \Pi(t,K_0) = \sum_{k>K_0} \sum_{m \le K_0} T_{m\to k}(t),
\end{equation}
where $K_0$ will be called spectral mode cutoff.
This quantity measures the overall tendency of the system to transfer activity toward higher or lower than $K_0$ spectral modes. Positive values of $\Pi(t,K_0)$ correspond to forward transfer, i.e. redistribution toward higher graph spectral modes. Negative values correspond to inverse transfer toward lower modes.
Importantly, this flux does not represent physical transport on the graph itself. Rather, it quantifies redistribution between structural scales defined by the graph Laplacian eigenbasis. The temporal evolution of $\Pi(t,K_0)$ therefore provides a direct characterisation of spectral cascade dynamics, as can be seen in Figure \ref{fig:ModalVariables}d, where forward cascade occurs.

\subsubsection{Persistence and Interaction Observables}

The final set of spectral observables introduced in this work characterises the global organisation of the transfer matrix.
As previously mentioned the transfer matrix can be classified into two parts by modes interaction. Diagonal part of transfer matrix is responsible for modal self-persistence. To quantify this process, we define the persistence intensity
\begin{equation}
    D(t) = \sum_m |T_{m\to m}(t)|.
\end{equation}
Similarly, we define the interaction intensity
\begin{equation}
   I(t) = \sum_{m\neq k} |T_{m\to k}(t)| ,
\end{equation}
which measures the total strength of inter-modal interactions.
From these quantities we construct the interaction ratio
\begin{equation}
   Q(t) = \frac{I(t)}{D(t)+I(t)} .
\end{equation}

This observable provides a normalized measure of the relative importance of interaction-mediated dynamics.
Specifically $Q(t)\approx 0$ corresponds to persistence-dominated dynamics, while $Q(t)\approx 1$ corresponds to interaction-dominated dynamics. The quantity Q(t) will be shown to reveal dynamical regimes that are largely invisible to conventional synchronisation measures.

\subsection{Conventional Synchronisation Measure}

For comparison with the spectral observables, we also monitor the standard Kuramoto synchronisation order parameter \cite{Kuramoto1984}
\begin{equation}
   R(t) = \left| \frac{1}{N} \sum_{j=1}^{N} e^{i\theta_j(t)} \right| .
\end{equation}

\begin{figure}[h]
\centering
\includegraphics[width=0.9\textwidth]{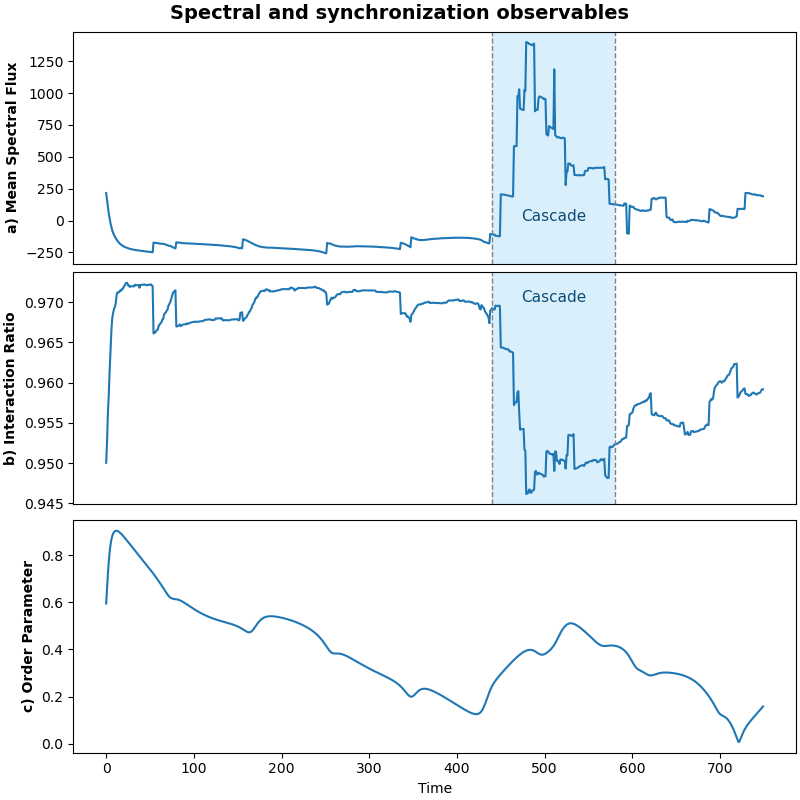}
\caption{Comparison of synchronisation and transfer observables for a representative network realisation. The figure consists of three time-dependent panels: (a) mean spectral flux $\Pi(t)$ averaged over the spectral cutoff parameter $K_0$, (b) interaction ratio $Q(t)$, and (c) synchronisation order parameter $R(t)$. The figure illustrates the relationship between graph-spectral transfer dynamics and global synchronisation. While the synchronisation order parameter remains comparatively smooth, the transfer observables exhibit pronounced fluctuations, intermittent bursts, and rapid reorganisations. These results demonstrate that substantial changes in graph-spectral transfer dynamics may occur even when the level of global synchronisation changes only weakly, highlighting the complementary information provided by transfer-based observables.}
\label{fig:SpectralObservables}
\end{figure}

The order parameter measures the degree of global phase coherence in the system. Values of $R$ close to unity indicate strong synchronisation, whereas values close to zero correspond to incoherent dynamics. One of the central observations of this work is that highly structured spectral transfer dynamics may persist even when $R(t)$ remains nearly stationary. An example is shown in Fig.~\ref{fig:SpectralObservables}, where the order parameter remains comparatively smooth despite pronounced changes in the spectral flux and transfer dynamics.

\subsection{Interpretation of Spectral Transfer Dynamics}

The framework introduced above allows synchronisation dynamics to be interpreted as a process of evolving spectral organisation.
Within this perspective:
\begin{itemize}
    \item graph Laplacian eigenmodes define structural dynamical scales,
    \item nonlinear coupling induces transfer between these scales,
    \item transfer matrices encode instantaneous interaction geometry,
    \item spectral flux characterises cascade directionality,
    \item interaction observables quantify the balance between persistence and redistribution.
\end{itemize}

This viewpoint extends conventional synchronisation analysis beyond global coherence and reveals a rich internal dynamics of spectral transfer episodes, intermittency, and cascade-like reorganisation hidden beneath apparently stable macroscopic synchronisation states.

\section{Computational Methodology}
Before investigating the transfer dynamics in detail, we describe the computational methodology. This section outlines the network construction procedure and the implementation of the Kuramoto dynamics.


\subsection{Network Construction}

The primary systems studied in this work are modular undirected graphs composed of densely connected communities linked by weaker inter-community connections. Such architectures are particularly suitable for graph-spectral analysis because low-order Laplacian modes naturally encode large-scale community organisation, whereas higher-order modes capture progressively finer structural detail. Synchronisation dynamics may therefore redistribute activity across well-defined graph scales.

Networks are generated using a stochastic block-model-type construction. This approach provides a controlled framework for studying transfer dynamics while retaining sufficient structural variability between network realisations. To ensure reproducibility, all simulations are performed using fixed random seeds.

Unless otherwise stated, the generated networks consist of:
\begin{itemize}
\item 5 modules,
\item module sizes sampled uniformly between 1 and 15 nodes,
\item intra-module connection probability $p_{\mathrm{in}} = 0.4$,
\item inter-module connection probability $p_{\mathrm{out}} = 0.03$.
\end{itemize}

These parameters generate networks with pronounced modular structure while preserving an element of random variability within and between communities. All networks considered in this study are undirected and unweighted.

Having specified the network construction procedure, we next describe the parameters governing the Kuramoto dynamics. The evolution of the oscillator phases is given by

\begin{equation}
\dot{\theta}*i = \omega_i + K \sum*{j=1}^{N} A_{ij}\sin(\theta_j-\theta_i),
\end{equation}

where:
\begin{itemize}
\item $\theta_i(t)$ denotes the phase of node $i$,
\item $\dot{\theta}*i$ denotes its time derivative,
\item $\omega_i$ is the natural frequency of node $i$,
\item $K$ is the global coupling strength,
\item $A*{ij}$ denotes the adjacency matrix of the network.
\end{itemize}

The network adjacency matrix $A$ is specified by the construction procedure described above. Initial phases are sampled independently from a uniform distribution, $\theta_i(0)\sim U(0,2\pi)$. The natural frequencies $\omega_i$ and the coupling strength $K$ constitute the principal control parameters governing both the synchronisation regime and the strength of nonlinear modal interactions. Their influence on the emergence of cascade-like transfer dynamics is investigated throughout this study.

To introduce intrinsic heterogeneity into the oscillator population, natural frequencies are sampled from a Gaussian distribution,

\begin{equation}
\omega_i \sim \mathcal{N}(0,\sigma_\omega^2),
\end{equation}

where $\sigma_\omega$ controls the degree of frequency heterogeneity.

Unless otherwise stated, the results presented in Figures~\ref{fig:ModalVariables} and \ref{fig:SpectralObservables} were obtained using $K=70$ and $\sigma_\omega=2.5$. These parameter values provide a balance between global synchronisation and the emergence of cascade-like transfer phenomena, as discussed in the following section.

\section{Dynamical Transfer Regimes}
The transfer matrix framework introduced above enables the investigation of dynamical processes that remain largely invisible to conventional synchronisation measures. While the Kuramoto order parameter quantifies the degree of collective phase coherence, it does not directly reveal how dynamical activity is redistributed among graph spectral modes. The transfer observables introduced in the previous sections provide access to this complementary aspect of the dynamics.
Our numerical experiments reveal that spectral transfer is not uniformly distributed throughout parameter space. Instead, the system exhibits several qualitatively distinct modes of organisation characterised by differing levels of synchronisation, transfer persistence, cascade activity, and reversal dynamics. These observations suggest the existence of dynamical transfer regimes that emerge from the interplay between coupling strength and frequency heterogeneity.
\subsection{Parameter-Space Organisation}
To explore the global structure of the dynamics, simulations were performed across a two-dimensional parameter space spanned by the coupling strength $K$ and the standard deviation of natural frequencies $\sigma_{\omega}$.
For each parameter pair $(K,\sigma_{\omega})$, the transfer matrix was computed and subsequently characterised through several derived observables including:
\begin{itemize}
    \item mean synchronisation level $\langle R(t) \rangle$,
    \item mean interaction ratio $\langle Q(t) \rangle$,
    \item spectral flux volatility $\sigma_\Pi$,
    \item burstiness measure $B$,
    \item forward occupancy fraction $P_{+}$,
    \item reversal rate $S$.    
\end{itemize}
Since the transfer matrix evolves in time and therefore forms a three-dimensional data structure, a set of derived observables is introduced to summarise its behaviour and reveal transitions across the parameter space.
First two quantities are means over time to observe general behaviour of Kuramoto order parameter and transfer matrix structure. Spectral flux volatility $\sigma_\Pi$ is defined as standard deviation of flux $\sigma_\Pi^2 = \left\langle (\Pi-\langle\Pi\rangle)^2 \right\rangle$ and burstiness measure $B = \sigma_\Pi / \langle |\Pi| \rangle$. Both those measure indicate how flux vary from its mean thus how likely cascade effects occur. 
Last two observables measure flux sign, that is if cascade is facing forward or backward. First to find a cascade, we create a threshold that cut random noise in flux equilibrium $P_c = 0.1 \langle |\Pi| \rangle$. Then for every value flux value we apply sign function
\begin{equation}
    \mathrm{sgn}(\Pi) = \left\{ \begin{array}{rcl}
+1 & \mbox{for}
& \Pi>P_c \\ 0 & \mbox{for} & -P_c \leq\Pi\leq P_c  \\
-1 & \mbox{for} & \Pi<-P_c
\end{array}\right.
.
\end{equation}
The forward occupancy fraction $P_{+}$ is defined as the fraction of spectral flux values that are positive. The reversal rate $S$ measures the frequency with which the flux changes sign between consecutive time steps.


\subsection{Overview of Dynamical Regimes}
\begin{figure}[h]
\centering
\includegraphics[width=0.9\textwidth]{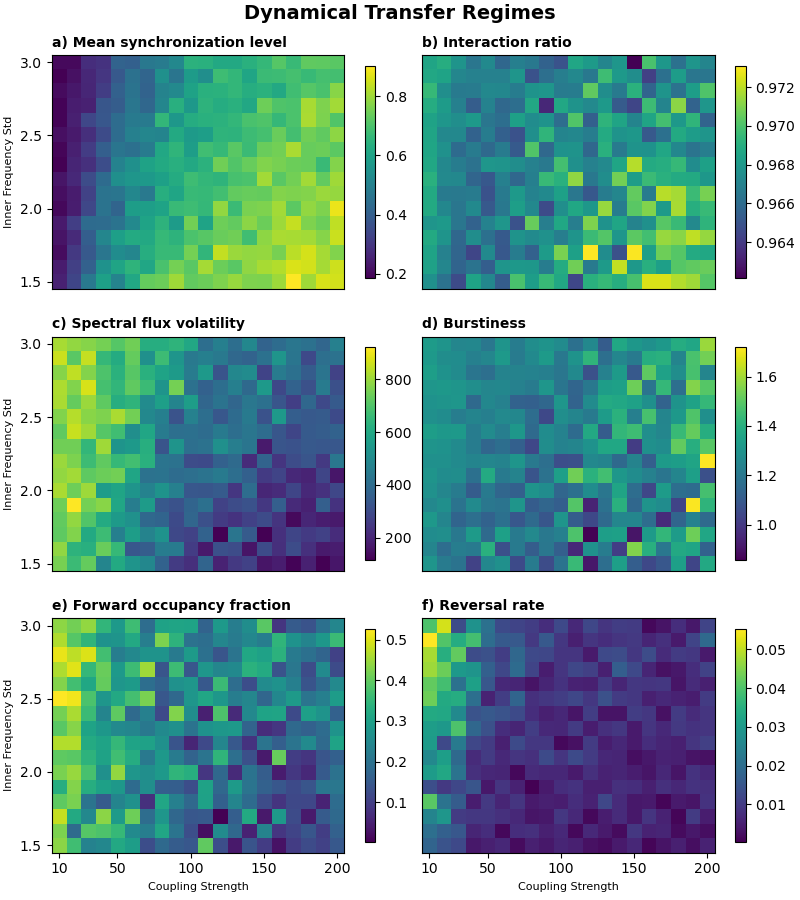}
\caption{Transfer dynamics in the parameter space spanned by the coupling strength $K$ and the standard deviation of natural frequencies $\sigma_{\omega}$. To reduce statistical fluctuations, all maps show averages over eight independently generated network realisations. Panels show: (a) mean synchronisation level $\langle R(t)\rangle$, (b) mean interaction ratio $\langle Q(t)\rangle$, (c) spectral flux standard deviation $\sigma_{\Pi}$, (d) flux burstiness measure $B$, (e) forward occupancy fraction $P_{+}$, and (f) reversal rate $S$. Together, these observables characterise synchronisation, transfer intensity, cascade variability, directional bias, and reversal dynamics across parameter space.
}
%
\label{fig:TransferRegimes}
\end{figure}

The heatmaps presented in Figure \ref{fig:TransferRegimes} reveal that the transfer observables are not distributed randomly throughout parameter space. Instead, coherent gradients emerge across the $(K,\sigma_{\omega})$ plane, indicating systematic changes in the organisation of the dynamics as coupling strength and frequency heterogeneity are varied.
A particularly clear example is provided by the synchronisation level and spectral flux volatility, which exhibit complementary transitions across parameter space. Regions of strong synchronisation are generally associated with reduced transfer fluctuations, whereas elevated transfer volatility tends to occur in less synchronised states. Similar trends are visible in the occupancy and reversal statistics, suggesting that the various observables capture different aspects of a common underlying dynamical organisation.
Taken together, the heatmap analysis and observable-space projections indicate that the system does not evolve continuously through a featureless parameter space. Instead, the dynamics cluster into a small number of qualitatively distinct regimes characterised by different combinations of synchronisation, transfer persistence, volatility, burstiness and reversal activity.
For clarity, we distinguish four representative regimes:
\begin{itemize}
\item \textbf{Regime I: Ordered State} -- characterised by strong synchronisation, weak transfer fluctuations and persistent spectral organisation.
\item \textbf{Regime II: Metastable State} -- characterised by substantial synchronisation coexisting with enhanced transfer activity, intermittent cascades and frequent spectral reorganisation.
\item \textbf{Regime III: Disordered State} -- characterised by weak synchronisation and fragmented transfer dynamics lacking long-lived coherent organisation.
\item \textbf{Regime IV: Mixed Transition Region} -- characterised by competing dynamical tendencies, increased variability and gradual transitions between the previous regimes.
\end{itemize}
The following subsections discuss these regimes in greater detail and relate their dynamical signatures to the underlying transfer processes revealed by the graph-spectral framework.

\subsection{Ordered State}
For sufficiently large coupling strengths and relatively small frequency heterogeneity, the system approaches a highly synchronised state characterised by large values of the order parameter.
In this regime, transfer matrices exhibit strong diagonal dominance, indicating that spectral modes primarily preserve their activity rather than exchange it with neighbouring modes. Spectral flux remains weak and relatively stable, while reversal events occur only rarely.
From the perspective of graph spectral dynamics, this regime corresponds to a state of persistent organisation in which large-scale coherent structures dominate the dynamics and suppress extensive redistribution between graph scales.

\subsection{Disordered State}
At the opposite end of parameter space, large frequency heterogeneity and weak coupling inhibit the formation of global synchronisation.
Transfer activity becomes fragmented and irregular, with reduced persistence of coherent spectral structures. Although local transfer events continue to occur, they fail to organise into sustained cascade-like patterns. Synchronisation levels remain low and transfer observables indicate the absence of long-lived large-scale organisation.
This regime may therefore be interpreted as a dynamically disordered state in which nonlinear interactions are insufficient to generate coherent collective behaviour.

\subsection{Metastable State}
Between the ordered and disordered limits, a broad intermediate region emerges in which transfer dynamics become particularly rich.
This regime is characterised by:
\begin{itemize}
    \item substantial synchronisation,
    \item enhanced transfer volatility,
    \item recurrent cascade reversals,
    \item intermittent bursts of transfer activity,
    \item strong off-diagonal transfer matrix structure.
\end{itemize}
A remarkable feature of this regime is that synchronisation may remain relatively stable while transfer observables fluctuate strongly. Consequently, states that appear similar when viewed through conventional order parameters can exhibit dramatically different internal transfer organisation.
The metastable regime therefore represents a region of persistent dynamical reorganisation, where graph spectral activity continuously redistributes among competing collective structures.

\subsection{Mixed Transition Region}
The boundaries separating the previous regimes are not sharp.
Instead, the system passes through broad transition zones characterised by strong variability across realisations and increased sensitivity to initial conditions.
Within these regions, transfer observables frequently display competing tendencies. For example, synchronisation may increase while reversal activity remains elevated, or forward cascade occupancy may coexist with substantial flux volatility. Such behaviour suggests the coexistence of multiple dynamical mechanisms operating on comparable timescales.
These mixed states occupy an important position within the overall phase diagram because they appear to mediate the transition between stable spectral organisation and strongly fluctuating transfer dynamics.

\subsection{Observable-Space Separation}
\begin{figure}[t]
\centering
\includegraphics[width=1.1\textwidth]{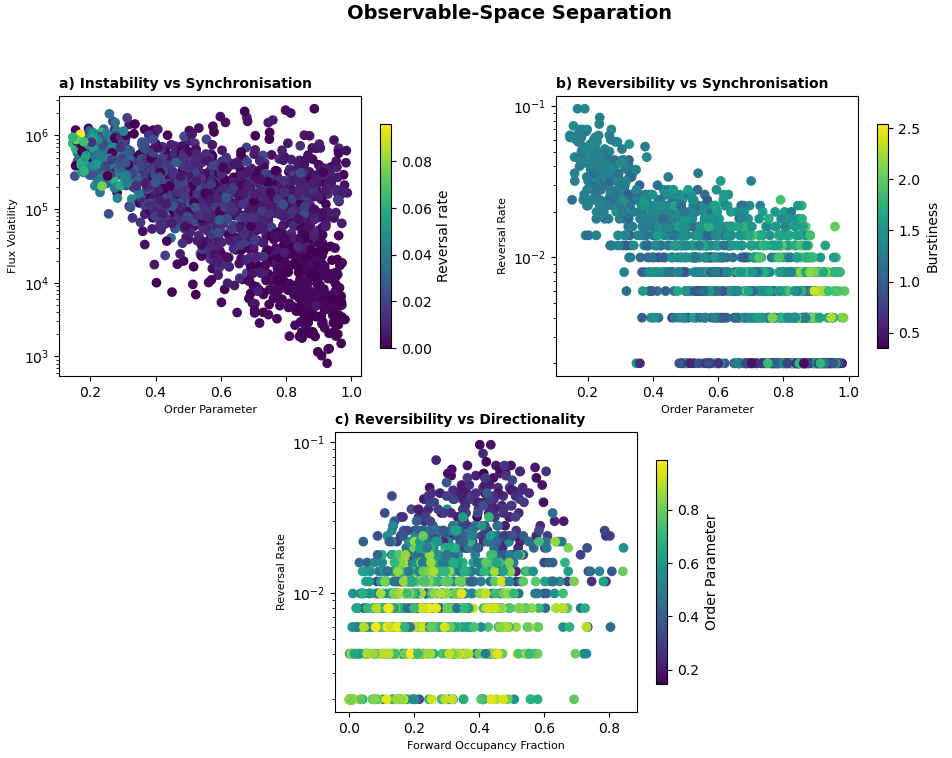}
\caption{Observable-space representation of transfer dynamics. Each point corresponds to a single parameter pair $(K,\sigma_{\omega})$. Distinct regions of the observable space correspond to different transfer regimes. Panels illustrate relationships between (a) order parameter $R$ and reversal rate $S$, (b) correlation of burstiness $B$ and reversal rate $S$, (c) connection between forward occupancy fraction $P_{+}$ and order parameter $R$.}
\label{fig:observable_space} 
\end{figure}

Additional insight can be obtained by examining the dynamics in the space of transfer observables rather than directly in parameter space.
Each parameter pair $(K,\sigma_{\omega})$ may be represented by a feature vector
\begin{equation}
\mathbf{x} = \left(\langle R \rangle, \sigma_\Pi,B,P_{+},S\right),
\end{equation}
which characterises the corresponding transfer state.
Scatter plots in this observable space, presented in Fig.~\ref{fig:observable_space}, reveal systematic relationships between synchronisation, transfer volatility, reversal activity and directional occupancy. In particular, increased reversal activity is generally associated with reduced synchronisation and enhanced transfer fluctuations, suggesting that the various observables capture complementary aspects of a common underlying organisation process.

Taken together, the heatmap analysis and observable-space projections provide evidence that spectral transfer dynamics are organised into a small number of reproducible dynamical regimes. These regimes are not defined solely by synchronisation level but instead emerge from the combined behaviour of transfer persistence, cascade directionality, volatility and reversal activity. The transfer framework therefore reveals an additional layer of dynamical organisation that remains largely hidden from conventional synchronisation analysis.

\section{Discussion}

The present work lies at the intersection of graph signal processing (GSP) and network synchronisation. Existing GSP approaches, such as spectral graph wavelet transforms, provide tools for analysing non-stationary graph signals and identifying time-localised spectral structures \cite{wavelet1,wavelet2}. However, these methods do not explicitly quantify interactions between graph-spectral modes. Conversely, synchronisation studies traditionally focus on global observables such as the Kuramoto order parameter and related measures of collective coherence \cite{Acebron2005,Arenas2008}. While spectral properties of the graph Laplacian are frequently used to characterise network structure and synchronisation pathways \cite{Timofeyev_2025}, relatively little attention has been paid to the transfer of dynamical activity between graph-spectral modes.

The present work introduces a graph-spectral framework for analysing collective dynamics in networks of coupled oscillators. Rather than focusing exclusively on synchronisation levels, the proposed approach characterises how dynamical activity is redistributed among Laplacian modes through nonlinear interactions. The resulting transfer matrix provides a time-resolved description of spectral interactions and enables the definition of transfer observables such as spectral flux, occupancy fractions, reversal rates and transfer volatility. Together, these quantities reveal aspects of the dynamics that are not directly accessible through conventional order parameters.

A particularly noteworthy observation is that substantial changes in transfer organisation may occur while the global synchronisation level remains relatively stable. In several parameter regions, the order parameter exhibits only modest variations whereas spectral flux and transfer observables display pronounced fluctuations, reversals and intermittent bursts. This suggests that synchronisation and transfer capture complementary dimensions of collective behaviour. In modular networks, the order parameter alone may not fully characterise the macroscopic dynamical state, since oscillators can organise into partially synchronised clusters.

From a broader perspective, the observed dynamics bear qualitative similarities to transfer processes known from multiscale physical systems. However, the cascades identified here do not occur in conventional Fourier space but rather in the graph spectral domain defined by the Laplacian eigenmodes of the network. The transfer matrix therefore provides a bridge between synchronisation dynamics and graph signal processing, enabling the study of redistribution processes across graph scales.
The parameter-space analysis further indicates that transfer dynamics are organised into distinct regimes. Ordered, metastable and disordered behaviours emerge naturally from the interplay between coupling strength and frequency heterogeneity. These regimes are more clearly distinguished by transfer observables than by synchronisation measures alone, suggesting that graph-spectral transfer may provide an additional layer of dynamical characterisation for networked systems.

The present study should be regarded as a first step towards a broader theory of spectral transfer on graphs. Several important questions remain open. These include the role of network topology, the influence of modularity and community structure, the dependence on graph size, and the generality of the observed transfer regimes across different classes of nonlinear dynamical systems. Future work will investigate these questions in greater detail.

Taken together, the results demonstrate that graph spectral transfer provides a useful framework for studying the internal organisation of collective dynamics. By revealing redistribution processes hidden beneath conventional synchronisation measures, the approach opens new possibilities for the analysis of complex networked systems.

\section{Conclusions}
In this work, we introduced a graph-spectral framework for analysing collective dynamics in networks of coupled oscillators. By combining graph Fourier decomposition with a transfer matrix formalism, we developed a methodology capable of quantifying nonlinear interactions between Laplacian modes and tracking the redistribution of dynamical activity across graph spectral scales.

The proposed framework extends conventional synchronisation analysis by providing access to internal transfer processes that remain largely hidden from global order parameters. While synchronisation measures quantify the degree of collective coherence, the transfer matrix reveals how activity is exchanged among collective graph modes and how this exchange evolves in time.
Applying the framework to modular networks of Kuramoto oscillators revealed a rich spectrum of transfer phenomena. In addition to persistent transfer states, we observed intermittent bursts, cascade reversals, and strongly fluctuating transfer episodes. These dynamics were quantified through a set of transfer observables including spectral flux, transfer volatility, occupancy fractions, reversal statistics, and burstiness measures.

A central finding of the present study is that transfer dynamics may exhibit substantial reorganisation even when the global synchronisation level changes only weakly. This indicates that synchronisation and spectral transfer represent complementary aspects of collective behaviour and suggests that important dynamical processes may remain undetected when analysis is restricted to conventional order parameters alone.

The parameter-space investigation further revealed the emergence of distinct dynamical transfer regimes organised by the interplay between coupling strength and frequency heterogeneity. Ordered, metastable, and disordered behaviours could be identified through characteristic combinations of synchronisation, transfer persistence, volatility, and reversal activity. These results suggest that graph-spectral transfer provides an additional layer of dynamical organisation beyond traditional synchronisation-based descriptions.
More broadly, the present work establishes a connection between synchronisation dynamics, graph signal processing, and transfer phenomena in complex systems. The observed cascades occur not in conventional Fourier space but in the graph spectral domain defined by network topology, thereby opening a new perspective on multiscale organisation in networked dynamical systems.

The framework introduced here represents a first step towards a broader theory of spectral transfer on graphs. Future investigations should address the role of network topology, graph size, modularity, and heterogeneous architectures, as well as the applicability of the transfer formalism to other classes of nonlinear dynamical systems. Such studies may help clarify the general principles governing the emergence, persistence, and reorganisation of collective dynamics in complex networks.
Taken together, the results demonstrate that graph spectral transfer constitutes a powerful and interpretable framework for studying collective dynamics. By revealing redistribution processes hidden beneath conventional synchronisation measures, it provides new tools for investigating the organisation of complex systems across graph scales.

\section*{Data Availability}
Code and simulation datasets are available on reasonable request through the corresponding author.

\bibliography{bibliography.bib}

\end{document}